\documentclass[12pt]{article}
\usepackage{amsfonts}
\usepackage{pslatex}
\usepackage{amssymb}
\usepackage{epsfig}
\usepackage{latexsym}
\input amssym.def
\input amssym.tex
\usepackage{graphicx}
\usepackage{color}
\usepackage{cite}

\begin{document}
\thispagestyle{empty}

\def\a{\alpha}
\def\b{\beta}
\def\g{\gamma}
\def\d{\delta}
\def\dd{\rm d}
\def\e{\epsilon}
\def\ve{\varepsilon}
\def\z{\zeta}

\def\B{\mbox{\bf B}}

\begin{flushright}
SNUST-071201\phantom{abc}\\
UOSTP-07107\phantom{abcd}\\
\end{flushright}
\vspace{.3cm}

\renewcommand{\thefootnote}{\fnsymbol{footnote}}
\centerline{\Large \bf Reflecting Magnon Bound States}

\vskip 1.2cm
\centerline{
Changrim  Ahn$^{1}$, \,\,
Dongsu Bak$\!^2$, \,\,
Soo-Jong Rey$^3$
}

\vskip 10mm
\renewcommand{\thefootnote}{\arabic{footnote}}
\setcounter{footnote}{0}

\centerline{ $^1$ \sl
Department of Physics, Ewha Womans University, Seoul 112-750 Korea}
\vskip 3mm
\centerline{$^2$ \sl
Physics Department, University of Seoul, Seoul 130-743 Korea}
\vskip 3mm

\centerline{$^3$ \sl School of Physics and Astronomy,
Seoul National University, Seoul 151-747 Korea
\phantom{abcdefg}}
\vspace*{0.6cm}
\centerline{\tt ahn@ewha.ac.kr, dsbak@mach.uos.ac.kr, sjrey@snu.ac.kr}

\vskip 20mm

\baselineskip 18pt

\begin{center}
{\bf Abstract}
\end{center}
\begin{quote}
In ${\cal N}=4$ super Yang-Mills spin chain, we compute reflection amplitudes
of magnon bound-state off giant graviton. We first compute the reflection amplitude off $Y=0$ brane boundary and compare it with the scattering amplitude between two magnon bound-states in the bulk. We find that analytic structure of the two amplitudes are intimately related each other: the boundary reflection amplitude is a square-root of the bulk scattering amplitude. Using such relation as a guide and taking known results at weak and strong coupling limits as inputs, we find the reflection amplitude of an elementary magnon off $Z=0$ giant graviton boundary. The reflection phase-factor is shown to solve crossing and unitarity relations. We then compute the reflection amplitude of magnon bound-state off the $Z=0$ brane boundary and observe that its analytic structures are
again intimately related to the bulk scattering and the $Y=0$ boundary reflection amplitudes. We also take dyonic giant magnon limit of these reflection amplitudes and
confirm that their phase-shifts agree completely with string worldsheet
computations based on complex sine-Gordon soliton scattering.

\end{quote}
\vskip 1cm
\centerline{\today}
\newpage
\baselineskip 18pt

\def\nn{\nonumber}
\def\tr{{\rm tr}\,}
\newcommand{\bea}{\begin{eqnarray}}
\newcommand{\eea}{\end{eqnarray}} \newcommand{\bde}{{\bf e}}
\renewcommand{\thefootnote}{\fnsymbol{footnote}}
\newcommand{\be}{\begin{equation}}
\newcommand{\ee}{\end{equation}}

\vskip 0cm

\section{Introduction}
The newly discovered integrable structure \cite{MZ} of the planar ${\cal N}=4$ super Yang-Mills theory played important role in testing the AdS/CFT correspondence \cite{AdS/CFT} over {\sl all} range of the `t Hooft coupling parameter. By mapping the dilatation operator to an integrable spin-chain, scaling dimension of single trace operators is computable to all orders in perturbation theory \cite{Beisert}. The spectrum is then compared with the excitation energy spectrum of a free closed string in AdS$_5 \times \mathbb{S}^5$ with large angular momenta. Important physical observables in this setup are the spectrum and the states. Worldsheet scattering $S$-matrices offer a powerful method for extracting them \cite{staudacher}. Utilizing underlying symmetries, Beisert \cite{BeisertSmatrix} derived the $S$-matrices up to an overall phase-factor. This phase-factor contains important dynamical information and was later determined by Beisert, Eden and Staudacher \cite{BES}. The phase-factor was shown to satisfy a certain crossing relation \cite{Janik}.

A new interesting feature arises upon introducing boundaries. In integrable quantum field theories, in the presence of boundaries, full integrability of the bulk can be maintained only for appropriate choices of boundary condition. The same situation arises in AdS/CFT correspondences \cite{boundaries}. In the string theory side, D-branes introduce the boundary to string worldsheet. In the ${\cal N}=4$ super Yang-Mills (SYM) side, bifundamental or subdeterminant field (products) introduce boundaries to composite operators. Not all boundaries would maintain integrability. Recently, Hofman and Maldacena \cite{HM} investigated two integrable boundary conditions which correspond to maximal configurations of a giant graviton interacting with elementary magnons of the spin chain
attached to it. There are two kinds of them. One is the $Y=0$ brane,
represented by composite operators containing a determinant factor ${\rm det}(Y)$:
\be
{\cal O}_Y=\epsilon^{j_1\ldots j_{N-1}A}_{i_1\ldots i_{N-1}B}
Y^{i_1}_{j_1} \cdots Y^{i_{N-1}}_{j_{N-1}}
(Z \ldots Z \chi Z\ldots Z\chi' Z\ldots)^B_A,
\label{ybrane}
\ee
where $\chi,\chi',\ldots$ represent other SYM fields. Another is $Z=0$ brane, represented by composite SYM operators containing a determinant factor ${\rm det}(Z)$:
\bea
{\cal O}_Z = \epsilon^{j_1 \cdots j_{N-1} A}_{i_1 \cdots i_{N-1} B}
Z_{j_1}^{i_1} \cdots Z_{j_{N-1}}^{i_{N-1}} (\chi Z \cdots Z \chi' Z \cdots Z\chi''Z \cdots \chi''')^B_A \,\,.\label{zbrane}
\eea
An important difference of $Z=0$ brane from the $Y=0$ brane is that the open super Yang-Mills spin chain is connected to the giant graviton through boundary impurities $\chi$ and $\chi'''$. In this paper, for simplicity, we shall take $\chi = \cdots = \chi''' = Y$.
The dilatation operator determining the conformal dimension of these operators has been derived and mapped to the integrable spin chain models with appropriate boundary conditions. The corresponding boundary $S$-matrices were obtained in \cite{HM} up to boundary dressing phase-factor. Recently, this factor was determined from boundary crossing relation by Chen and Correa \cite{CC} for $Y=0$ brane. On the other hand, the corresponding factor for $Z=0$ brane is unknown.
With the boundary terms preserving integrability, this system can be
completely described by the reflection scattering matrix (namely, boundary $S$-matrix) which preserves particle numbers and energies in the same way as the bulk scattering matrix does.
On the other hand, momenta are reversed.

In this paper, using fusion procedure, we construct complete set of reflection amplitudes of magnons and their bound-states off a giant graviton and compare analytic structure of these amplitudes with that of bulk scattering amplitudes between magnon bound-states. Thus, in section 2, we first recapitulate relevant results of the bulk scattering amplitudes. In section 3, utilizing the boundary dressing phase-factor of \cite{CC}, we study reflection amplitudes of a magnon bound-state off the $Y=0$ brane. We find a remarkable structure that the reflection amplitude takes a square-root form of the bulk scattering amplitude. Taking this relation as a guide and utilizing known strong and weak coupling results \cite{HM}, we then study in section 4 the $Z=0$ brane as well. We first find the reflection dressing phase-factor for an elementary magnon and show that it satisfies the crossing and the unitary conditions. Using it, we proceed to compute the reflection amplitude of a magnon bound-state off the $Z=0$ brane. We again confirm that the amplitude takes a square-root form of the bulk scattering amplitude that
involves a magnon bound-state {\sl and} boundary modes. From these amplitudes, we also extract the reflection phase-shifts of the dyonic giant magnon off both types of the giant gravitons. At strong coupling, the result may be compared with string theory worldsheet computations. In the latter, the phase-shift is computable from soliton scattering in complex sine-Gordon model. In section 4, we  compute these two results and find perfect agreement.

\section{Bulk S-Matrix of Magnon Bound-State}

The magnon bound-states \cite{magnonboundstate} constitute an important set of BPS excitations of a single closed string. Starting from the Bethe equation, scattering amplitudes between two magnon bound-states of charge $m$ and $n$ were constructed \cite{boundstateSmatrix}. The same result is also obtainable \cite{ours} from Beisert's $S$-matrices \cite{BeisertSmatrix}. Consider two magnon bound-states $\B^{(m)}, \B^{(n)}$
\bea
\B^{(m)} &=& |Y_1 \cdots Y_m\rangle \ \leftrightarrow \
{\rm tr}( Z \cdots Z Y_1 Z \cdots Z \cdots Y_m Z \cdots Z)
\nonumber \\
\B^{(n)} &=&
 |Y_1 \cdots \, Y_n\rangle \ \leftrightarrow \
{\rm tr}( Z \cdots Z Y_1 Z \cdots Z \cdots Y_n Z \cdots Z)
\label{magnonboundstate} \eea
formed by a complex adjoint scalar field $Y$ in the ferromagnetic ground-states of $Z$. We denote by $x^\pm_k$ the spectral parameters of elementary magnon inside $\B^{(m)}$:
\bea
x^\pm_k = {e^{\pm i p_k/2} \over
4g \sin {p_k \over 2}} \left( 1 + \sqrt{ 1 +
16g^2 \sin^2 {p_k \over 2}} \right)
\eea
and similarly by $y^\pm_k$ the spectral parameters of elementary magnons inside $\B^{(n)}$.
Here, $g^2 = g^2_{\rm YM} N_c / 16 \pi^2$. They obey the so-called multiplet shortening conditions \cite{BeisertSmatrix}:
\bea
x^+_k + {1 \over x^+_k}  - x^-_k - {1 \over x^-_k} = {i\over g}, \qquad
y^+_k + {1 \over y^+_k} - y^-_k - {1 \over y^-_k} = {i\over g}\,.
\eea
The elementary magnon has dispersion relation
\bea
E = \sqrt{ 1 + 16 g^2 \sin^2 \left({p \over 2} \right) }. \eea
In order for these elementary magnons to form bound-states, the spectral parameters ought to obey \cite{magnonboundstate, ours}
\bea
&& x^-_1 = x^+_2, \quad x^-_2 = x^+_3, \quad \cdots \quad x^-_{m-1} = x^+_m
\nonumber \\
&& y^-_1 = y^+_2, \quad y^-_2 = y^+_3, \quad \cdots \quad y^-_{n-1} = y^+_n
\eea
The spectral parameters of the bound-states $\B^{(m)}$ and $\B^{(n)}$ are given by
\bea
X^+ = x^+_1, \quad X^- = x^-_m; \qquad Y^+ = y^+_1, \quad Y^- = y^-_n
\eea
and obey the multiplet shortening conditions
\bea
X^+ + {1 \over X^+} - X^- - {1 \over X^-} = {m i\over g} ; \qquad
Y^+ +{1 \over Y^+} - Y^- - {1 \over Y^-} = {n i\over g}. \label{boundstatespecpara}\eea
The bound-states of charge $Q$ (which equals to $m, n$ in the present case) obey the dispersion relation
\bea
E_Q = \sqrt{Q^2 + 16 g^2 \sin^2 \left({p \over 2} \right)} \qquad \mbox{where} \qquad
e^{ip} = {X^+ \over X^-}. \eea
We are especially interested in analytic structure of scattering amplitudes. We thus begin with recapitulation of the structure for the bulk $S$-matrix of magnon bound-states.

We first recall how the $S$-matrix is computed. In the ferromagnetic vacuum, excitations are organized by chiral and antichiral supergroups $\mathfrak{psu}(2|2) \otimes \overline{\mathfrak{psu}(2|2)} \ltimes \mathbb{Z}^{2,1}$, extended by diagonal off-shell $sl(2)$ central charges. The physical excitations $(8 \vert 8)$ transform under each $\mathfrak{psu}(2\vert 2) \ltimes \mathbb{Z}^{2,1}$ as $(2 \vert 2)$ irreducibly. Overall, $(8 \vert 8) = (2 \vert 2) \otimes \overline{(2 \vert 2)}$. Since the centrally extended supergroup symmetries are identical, the ${\cal N}=4$ super Yang-Mills $S$-matrices are computed from product of chiral and antichiral $S$-matrices as
\bea
{\cal S}^{{\cal N}=4}_{ab}(x^\pm_a, x^\pm_b) = S_0(x^\pm_a, x^\pm_b) {\cal S}_{ab} (x^\pm_a, x^\pm_b)
{1 \over A (x^\pm_a, x^\pm_b)} \overline{{\cal S}_{ab}} (x^\pm_a, x^\pm_b)
\eea
Here, $A$ refers to the S-matrix of the highest state
\bea
A(x^\pm, y^\pm) = \frac{(x^- - y^+)}{(x^+ - y^-)} \label{A}
\eea
and $S_0$ is an overall phase-factor \cite{AF}:
\bea
S_0(x^\pm, y^\pm)=
\frac{(1-\frac{1}{x^- y^+})}{(1-\frac{1}{x^+ y^-})}
\frac{1}{\sigma^2(x^\pm, y^\pm)}\, . \label{S0} \eea
The dressing phase-factor $\sigma^2(x^\pm, y^\pm)$, introduced first in \cite{AFS}, is given by exponential of symplectic form of higher conserved charges \cite{BES}:
\bea
\sigma^2(x^\pm, y^\pm) = {1 \over \sigma^{2} (y^\pm, x^\pm)} = \frac{R^2(x^+, y^+)R^2(x^-, y^-)}
{R^2(x^+, y^-) R^2(x^-, y^+)}. \label{sigmasquare}
\eea
In the foregoing discussions, we do not need explicit expression for $R^2(x,y)$; the expression can be
found, for example, in \cite{DHM}.
For the highest state, $A(x^\pm, y^\pm)$ and the first factor in $S_0(x^\pm, y^\pm)$ combine into the Beisert-Dipple-Staudacher (BDS) $S$-matrix \cite{BDS}. We shall refer the first factor in (\ref{S0}) as BDS conversion factor.

As mentioned above, we restrict excitations to the scalar field $Y  \equiv \phi \overline{\phi}$. This simplifies the $S$-matrix computation considerably. The $S$-matrix is simply $A(x^\pm,y^\pm)$ in (\ref{A}), so the full scattering amplitude is essentially the same as $S_0(x^\pm,y^\pm)$ times $A(x^\pm, y^\pm)$ computed from the $\mathfrak{psu}(2|2) \ltimes \mathbb{Z}^{2,1}$ chiral supergroup:
\bea
{\cal S} \vert \phi(x) \phi(y) \rangle = S_0 (x^\pm, y^\pm) \, A(x^\pm, y^\pm) \vert \phi(y) \phi(x) \rangle. \label{bulkelementary}\eea
\begin{figure}[ht!]
\centering \epsfysize=9cm
\includegraphics[scale=0.7]{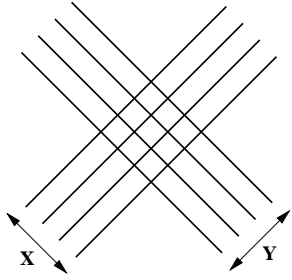}
\caption{\small \sl The bulk scattering between two magnon bound-states with spectral parameters $X^\pm, Y^\pm$. The BDS scattering amplitude originates from the diagonal vertex interactions. The extra phase-factor $\mathfrak{M}_{\B}$ originates from the off-diagonal vertex interactions.}
\end{figure}
Then, the 2-body ${\cal S}$-matrix between the magnon bound-states $\B^{(m)}, \B^{(n)}$ is computable by fusion procedure, as depicted in Fig. 1. The result is
\bea
{\cal S}\vert \B^{(m)}(X) \B^{(n)}(Y) \rangle = S_{\B} (X^\pm, Y^\pm) A (X^\pm, Y^\pm) \vert \B^{(n)}(Y) \B^{(m)}(X) \rangle \, . \label{bulkboundstate}
\eea
It takes exactly the same form as the elementary magnon scattering amplitude (\ref{bulkelementary}). So, $A (X, Y)$ is the $S$-matrix in (\ref{A}) except that the spectral parameters are now replaced by those of the bound-state (\ref{boundstatespecpara}). In fusion procedure, product of diagonal $S$-matrices in Fig. 1 gives rise to the BDS scattering matrix (viz. $A(X, Y)$ and the BDS conversion factor).
The factor $S_{\B}(X, Y)$ denote an overall phase-factor arising from product of off-diagonal $S$-matrices in Fig. 1:
\bea
S_{\B} (X^\pm, Y^\pm) = S_0 (X^\pm, Y^\pm) \mathfrak{M}_{\B} (X^\pm,Y^\pm)
= {(1 - {1 \over X^- Y^+}) \over (1 - {1 \over X^+ Y^-})} {\mathfrak{M}_{\bf B}(X^\pm, Y^\pm) \over \sigma^2(X^\pm, Y^\pm)} \, .
\label{bulkSB}
\eea
We shall make use of these anatomical observations when drawing a physical picture of boundary reflection amplitudes in the next sections.
In (\ref{bulkSB}), $S_0(X^\pm, Y^\pm)$ is the dressing phase-factor that appeared in the elementary scattering $S$-matrices in (\ref{S0}) except that the spectral parameters
are now replaced by those of the bound-state $X^\pm, Y^\pm$. The extra contribution $\mathfrak{M}_{\B} (X^\pm, Y^\pm)$ is the phase-factor that arises from the scattering amplitudes among the constituent magnons inside each bound-states.
For $m \le n$,
\bea \mathfrak{M}_{\B} (X^\pm, Y^\pm) = \left( {X^+ + {1 \over X^+} - Y^+ - {1
\over Y^+} \over X^- + {1 \over X^-} - Y^- - {1 \over Y^-}} \right)
\prod_{k=1}^{m-1} \left( - {X^+ +{1 \over X^+} - Y^+ - {1 \over Y^+} -
\frac{i k}{g} \over X^- + {1 \over X^-} - Y^- -{1 \over Y^-} +
\frac{i k}{g}} \right)^2. \label{boundstatedressing}\eea
The first part in the product represents the would-be $t$-channel pole. Notice that, by charge conservation of the scalar field $\Phi$ obeyed throughout the interactions, this part disappears when $m = n$.

In the strong coupling limit, the phase-factor (\ref{boundstatedressing}) features interesting analyticity properties as a function of the spectral variables. In the Hofman-Maldacena regime \cite{GiantMagnon} ($m, n$ held fixed as $g \rightarrow \infty$), the dressing phase-factor $S_0 (X^\pm, Y^\pm)$ dominates over $\mathfrak{M}(X^\pm, Y^\pm)$. In the dyonic giant magnon regime \cite{dorey} (the  `magnon density' $m/g, n/g$ held fixed as $g \rightarrow \infty$), $S_{\B}(X^\pm, Y^\pm)$ and $\mathfrak{M}(X^\pm, Y^\pm)$ are of the same order. This demonstrates that, at least in the strong coupling regime, functional form of the overall phase-factor $S_{\B}(X^\pm, Y^\pm)$ depends on the density of the elementary magnons only and not on other details of the bound-states. Therefore, we propose to take magnon bound-state as an interesting probe for diagnosing analytic structure of phase-factors that may also show up in other processes such as reflection scattering off a boundary.

\section{Reflection Amplitudes off \, $Y=0$ \, Brane}

With the motivations explained in the previous section, we now consider giant gravitons and scattering a magnon bound-state off them. The giant gravitons are BPS states in ${\cal N}=4$ super Yang-Mills theory and creates open boundary to the spin chain. In the AdS/CFT dual description, the giant gravitons are where open fundamental string ends. Schematically, the scattering between magnon bound-states and the scattering of magnon bound-state off the giant graviton are shown in
Fig. 1 and Fig. 2.

We take analytic structure of the bulk phase-factor (\ref{boundstatedressing}) as a useful guide for boundary reflection amplitudes. We shall be computing the boundary scattering ${\cal R}$-matrix explicitly for $Y=0$ and $Z=0$ branes and investigate the boundary phase-factors. For ${\cal R}$-matrix off the $Y=0$ brane, we shall find that the resulting boundary phase-factor is given in the form remarkably consistent with the bulk phase-factor (\ref{boundstatedressing}). Proceeding to the $Z=0$ brane, we shall motivate ourselves by taking these organizing structure of the dressing phase-factor as a guideline. We then put forward a proposal for the boundary phase-factor by taking account of all known results at both the weak and the strong coupling regimes. Our proposal takes a remarkably simple functional form, satisfies all consistency conditions and fully agrees with the aforementioned analytic structure of the bound-state phase-factor.
\begin{figure}[ht!]
\centering \epsfysize=9cm
\includegraphics[scale=0.7]{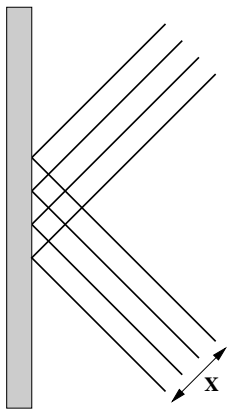}
\caption{\small \sl The reflection of magnon bound-state with spectral parameter $X^\pm$ off the left boundary. Reflection-double of the process across the boundary is related to the bulk scattering in Fig. 1.}
\end{figure}

For the case of $Y=0$ brane, the boundary breaks the excitation symmetry supergroup to
$\mathfrak{psu}(1\vert 2) \otimes \overline{\mathfrak{psu}(1 \vert 2)}$. The reflection matrix is given by
 \be
R^Y_{L}(x^{\pm})=R^Y_{0L}(x^{\pm})\left(
\begin{array}{cccc}
-\frac{x^+}{x^-}&0&0&0\\ 0&1&0&0\\ 0&0&1&0\\ 0&0&0&1
\end{array}\right),\quad
R^Y_{R}(x^{\pm})=R^Y_{0R}(x^{\pm})\left(
\begin{array}{cccc}
-\frac{x^-}{x^+}&0&0&0\\ 0&1&0&0\\ 0&0&1&0\\ 0&0&0&1
\end{array}\right).
\label{RLmatrix} \ee
Here, $R^Y_{0L}, R^Y_{0R}$ denote the corresponding reflection phase-factors.
As for the bulk, the full super Yang-Mills reflection matrix ${\cal R}^Y$ is computed by direct product of
$\mathfrak{psu}(1 \vert 2) \otimes \overline{\mathfrak{psu}(1 \vert 2)}$ chiral and antichiral $S$-matrices. For the reflection off either boundary, it is defined by
\bea
{\cal R}^Y (x^\pm) = R^Y_{0}(x^\pm) \, {R}^Y(x^\pm) {1 \over \widetilde{A}^Y(x^\pm)}
\overline{R}^Y(x^\pm). \eea
Here,
$\widetilde{A}^Y_L(x), \widetilde{A}^{Y}_R(x)$ are the reflection amplitudes of the highest state:
\bea
\widetilde{A}^{Y}_L (x^\pm) = - {x^+ \over x^-} \qquad \mbox{and} \qquad \widetilde{A}^{Y}_R (x^\pm) = - {x^- \over x^+}. \label{YbraneA}
\eea
The reflections off the left- and right-boundary are related by parity operation $P: x^\pm \rightarrow - x^\mp$.

The boundary phase-factors $R^Y_{0L}$ and $R^Y_{0R}$ must obey boundary crossing relations \cite{HM}:
\bea
R^Y_{0L}(x^{\pm})R^Y_{0L}({\bar x}^{\pm}) = {1 \over R^Y_{0R}(x^{\pm})R^Y_{0R}({\bar x}^{\pm})}
=\frac{\frac{1}{x^-}+x^{-}}{\frac{1}{x^+}+x^{+}}
\frac{1}{S_0(-{\bar x}^{\mp},x^{\pm})}
\label{Bcrossing} \eea
where
${\bar x}^{\pm}=1/x^{\pm}$ and $S_0$ is the overall phase-factor for bulk scattering given in (\ref{S0}). More recently, the boundary crossing relation (\ref{Bcrossing}) was
solved for the $Y=0$ brane \cite{CC}. As the BDS conversion factor in (\ref{S0}) becomes trivial in $S_0(x^\pm, -{\bar x}^\mp)$, the solutions for left- and right-boundary reflection are simply
\be R^Y_{0L}(x^{\pm})=\frac{x^-}{x^+}\sigma(x^{\pm},-x^{\mp}) \qquad \mbox{and} \qquad
R^Y_{0R}(x^{\pm})=\frac{x^+}{x^-}\sigma(-x^{\mp},x^{\pm}).
\label{Relementary} \ee
Taking them into account, the reflection amplitudes for the $Y= \phi \overline{\phi}$ magnon (which is the singlet under $\mathfrak{psu}(1 \vert 2) \otimes \overline{\mathfrak{psu}(1 \vert 2)}$) is given by
\bea
{\cal R}^Y_L \vert \phi(x^\pm) \rangle &=& R^{Y}_{0L}(x^{\pm}) \widetilde{A}^{Y}_L (x^\pm) \vert \phi(-x^\mp) \rangle
\nonumber \\
{\cal R}^Y_R \vert \phi(x^\pm) \rangle &=& R^Y_{0R}(x^{\pm}) \widetilde{A}^{Y}_R (x^\pm) \vert \phi(-x^\mp) \rangle \, .
\label{Y0elementary}\eea

We now consider scattering of the magnon bound-state $\B^{(n)}(X^{\pm})$ in (\ref{magnonboundstate}) off the $Y=0$ brane. As depicted in Fig. 2, the boundary reflection amplitude is computable via the fusion procedure. The result is
\bea
{\cal R}^Y_L \vert {\bf B}^{(n)}(X^\pm) \rangle = R^Y_{{\bf B} L} (X^\pm) \widetilde{A}^Y_L (X^\pm) \vert {\bf B}^{(n)}(-X^\mp) \rangle
\eea
and hence takes the same form as the elementary magnon amplitude.
Here, $\widetilde{A}^Y_L(X^\pm)$ is the left-reflection amplitude in (\ref{YbraneA}) except the spectral parameters refer to those of the bound-state ${\bf B}^{(n)}$. The boundary
phase-factor
\bea
R^Y_{{\bf B} L} (X^\pm) =
\widetilde{A}^Y_L (X^\pm) R^Y_{0L}(X^{\pm}) \mathfrak{M}^{Y}(X^\pm)
= - \sigma(X^\pm, - X^\mp) \mathfrak{M}^Y(X^\pm) \label{Y0amplitude}\eea
contains, much the same as the bulk scattering case, the bound-state phase-factor
\bea
\mathfrak{M}^{Y}(X^\pm) = \prod_{k=1}^{n-1}\left(-\frac{X^+ +\frac{1}{X^+}
-\frac{ik}{2g}}{X^-+\frac{1}{X^-} +\frac{ik}{2g}}\right).
\label{yzerobound} \eea
%
Remarkably, the reflection amplitude in (\ref{Y0amplitude}) is exactly the square-root of the bulk counterpart in (\ref{bulkboundstate}) upon taking $m=n$ and $Y^\pm = - X^\mp$ in the latter. In fusion procedure, this is evident from the observation that Fig. 1 is the same as reflection-double of Fig. 2 across the $Y=0$ boundary. As there is no localized mode at the boundary, in the reflection-double process, product of diagonal $S$ matrices ought to be absent. This means we should remove diagonal $A(X, Y)$ amplitude and the BDS conversion factor from the bulk scattering (\ref{bulkboundstate}) and identify square-root of the remaining product of off-diagonal $S$ matrices with the process in Fig. 2. This yields precisely (\ref{Y0amplitude}).
Recall that the would-be $t$-channel pole in (\ref{boundstatedressing}) disappears once $m=n$ is set for the present situation.

Contrary to the bulk factor which contains Coleman-Thun \cite{CT} type double poles, this boundary factor has simple poles. One might be tempted to interpret them as boundary bound-states. However, this is not the case: it is straightforward to check that these poles do not satisfy the boundary Bethe-Yang equations. Therefore they have nothing to do with formation of boundary bound-states. This fits with the fact that $Y=0$ brane does not support localized mode at the boundary. This also fits with the aforementioned relation for magnon bound-state scattering amplitudes that the boundary phase-factor should be viewed as square root of the bulk phase-factor.

\section{Reflection Amplitudes off \, $Z=0$ \, Brane}

We next compute reflection amplitude off the $Z=0$ giant graviton. Unlike the $Y=0$ brane case, there now exists a localized degree sitting at each boundary (as seen from the corresponding SYM operators in (\ref{zbrane})). Its spectral parameter is given by \cite{HM}
\be
x_B = {i\over 4g} (2+ \sqrt{2^2 + 16g^2})
\ee
with the relation
\be
x_B +{1\over x_B}={i\over g} \qquad \mbox{viz.} \qquad
x_B^+ + {1 \over x_B^+} - x_B^- - {1 \over x_B^-} = {2 i \over g}.
\ee
Notice that we expressed these relations in suggestive forms that the localized mode may be viewed as $n=2$ magnon bound-state at maximum momentum $p=\pi$ by interpreting $x_B^\pm \equiv \pm x_B$ in (\ref{boundstatespecpara}). Below, we shall find further supporting evidence of such interpretation. Presence of the boundary retains the full $\mathfrak{psu}(2|2) \otimes \overline{\mathfrak{psu}(2|2)}$ symmetry group. The boundary mode transforms as the fundamental representation under these groups. Their energy is again given by the central charge:
\be
E_B = {g\over i}\left(x_B -{1\over x_B}\right) = \sqrt{1 + 4 g^2}\,.
\ee

As for the bulk, the reflection matrix can be completely determined up to an overall phase-factor by utilizing the $\mathfrak{psu}(2|2)$ symmetry \cite{HM}. Here we consider the same type of scalar for both bulk ($\phi(x^\pm)$) and boundary ($\phi_B(x_B)$). We also focus on scattering off the left boundary. The right boundary result is obtainable by parity transformation $P: x^\pm \rightarrow - x^\mp$. Again, we define the full reflection matrix ${\cal R}^Z$ off the left boundary by
\bea
{\cal R}^Z_L (x, x_B) = R^Z_{0L} (x, x_B) \, R^Z_L (x, x_B) {1 \over \widetilde{A}^Z_L (x, x_B)} \overline{R^Z_L} (x, x_B),
\eea
where $R^{Z}_{0L}(x, x_B)$ denotes a reflection phase-factor
and $\widetilde{A}^Z_L (x,x_B)$ is the elementary reflection amplitude for the highest state
\be
\widetilde{A}_L^{Z} (x^\pm,x_B)= -{x^+\over x^-}\left({x^+ + x_B\over x^- - x_B} \right) =-{x^+ \over x^-} A (x_B^\pm, x^\pm).
\label{su22}
\ee
Thus, the full reflection amplitude for the scalar $Y = \phi \overline{\phi}$ is given by
\be
{\cal R}^Z_L|\phi_B(x_B) \phi(x^\pm)\rangle =
R^{Z}_{0L}(x^\pm, x_B) \widetilde{A}^{Z}_L(x^\pm,x_B) |\phi_B(x_B) \phi(-x^\mp)\rangle
\label{Zscattering} \ee
In (\ref{su22}), the first part originates from magnon reflection off the boundary and is the same as  $Y=0$ reflection amplitude. The second part depends on $x_B$, so it arises from  magnon scattering with the localized mode at the boundary. The last expression in (\ref{su22})
again supports the proposed interpretation of the localized mode as an $n=2$ magnon bound-state at maximum momentum $x^\pm = \pm x_B$.

We first determine the overall phase-factor $R_{0L}^{Z}$. Based on our result for
$Y=0$ brane and lower order result at strong and weak coupling limits of the $R_{0L}^{Z}$ presented in \cite{HM}, here we assert that the overall phase-factor is given by
\be
R^{Z}_{0L}(x,x_B)= {x^-\over x^+} \,\,
\left({x^+ + {1\over x^+}\over x^- + {1\over x^-} }\right)
\left({1 + {1\over x^+ x_B}\over 1 -{1\over x^- x_B}} \right)
\cdot \sigma(x^\pm, - x^\mp) \sigma^2(x^\pm,x^\pm_B). \label{ourproposal}
\ee
The first part encodes weak coupling perturbative results up to two loops, while the second part expressed in terms of dressing phase-factors encodes the strong coupling leading order results extracted from time-delay in sine-Gordon soliton scattering. We now argue that (\ref{ourproposal}) satisfies all requisite conditions.

First, (\ref{ourproposal}) is the minimal extension of the $Y=0$ brane to a situation a localized mode is present at the boundary. This is most transparently seen by arranging the scattering amplitude (\ref{Zscattering}) as
\be
R^{Z}_{0L}(x,x_B) \widetilde{A}^{Z}_L (x,x_B) = R^{Y}_{0L}(x) \widetilde{A}^{Y}_L (x)
\cdot S_{\B} (x_B, x) A (x_B, x). \label{decomposition}
\ee
In the right hand side, the first part originates from an elementary magnon scattering off empty ($Y=0$) boundary. The second
part is due to the localized mode: following the proposed interpretation of the localized mode
as $m=2$ magnon bound-state, it originates from bulk scattering between $m=2$ magnon bound-state (at maximal momentum $p=\pi$) and $n=1$ elementary magnon with $X = x_B, Y= x$ in (\ref{bulkSB}). The bulk scattering
amplitude in this case is given by
\be
S_{0} (x_B, x)= {1+{1\over x^+ x_B} \over 1-{1\over x^- x_B }}
 \sigma^2(x^\pm,x^\pm_B); \ \ \ \
 \mathfrak{M}_{\B}(x_B, x)
= \left( {x^- + {1 \over x^-} + x_B + {1 \over x_B} \over
 x^+ + {1 \over x^+} - x_B - {1 \over x_B} } \right)
= {x^+ + {1\over x^+}\over x^- + {1\over x^-} }. \label{Z0-SB}
\ee
Multiplying them, we find that they yield all the $x_B$-dependent parts in (\ref{ourproposal}) and (\ref{decomposition}).

Second, the proposed reflection phase-factor solves the crossing relation. Chiral $\mathfrak{psu}(2 \vert 2)$ part of the crossing relation was computed in \cite{CC}. Putting together both chiral and antichiral parts, we obtain the full $\mathfrak{psu}(2 \vert 2) \otimes
\overline{\mathfrak{psu}(2\vert 2)}$ crossing relation as
\bea
R^Z_{0L}(x^\pm) R^Z_{0L}(\bar{x}^\mp) = {1 \over R^Z_{0R} (x^\pm) R^Z_{0R} (\bar{x}^\mp)}
= \left( {x^- + {1 \over x^-} \over x^+ + {1 \over x^+}} \right) \sigma^2 (- \bar{x}^\mp, x^\pm)
\cdot h_B^2 \, \left( {x^+ + x_B \over x^- - x_B} \right) \left( {{1 \over x^+} + x_B \over
{1 \over x^-} - x_B} \right). \label{Zcrossingrelation}\eea
Here,
\bea
h_B(x^\pm, x_B) &=& {x^+ \over x^-} \left( {x^- - x_B \over x^+ - x_B} \right) { 1 + (x^+ x^- x_B)^2 \over ( 1 - x^+ x^-) (1 - (x^+ x_B)^2)} \nonumber \\
&=& \left( {x^- - x_B \over x^+ - x_B} \right) \left( {{1 \over x^-} + x_B \over {1 \over x^+} + x_B} \right), \eea
where the second line is obtained from the first by using the multiplet shortening conditions for $x^\pm$ and $x_B$. Compared to $Y=0$ brane case, extra part in the crossing relation (\ref{Zcrossingrelation}) arises from magnon scattering with the localized states at the boundary.
It is precisely accomodated by the $x_B$-dependent part in our proposed solution (\ref{ourproposal}). By a straightforward computation, we checked that our proposed phase-factor
(\ref{ourproposal}) solves the crossing relation (\ref{Zcrossingrelation}).

Third, the phase-factor (\ref{ourproposal}) satisfies the unitarity condition:
\bea
R^Z_{0L} (x^\pm, x_B) R^Z_{0L} (-x^\mp, x_B) = 1, \eea
provided reversed ordering in properly taken into account in scattering process between the magnon and the localized mode.

To elucidate our proposed reflection phase-factor (\ref{ourproposal}), we compute reflection
amplitude of the magnon bound-state ${\bf B}^{(n)}$ off the $Z=0$ brane, again using the fusion
method. For the left boundary, the result takes the same form as the elementary amplitude (\ref{Zscattering}):
\be
{\cal R}_L^Z|\phi_B(x_B) \B^{(n)} (X^\pm)\rangle =
R^Z_{{\bf B}L} (X, x_B) \widetilde{A}^{Z}_L (X,x_B)
|\phi_B(x_B) \B^{(n)} (-X^\mp)\rangle\,. \label{Z0-boundstateR}
\ee
Here,  $R^Z_{{\bf B}L}$ is the bound-state reflection phase-factor:
\bea
R^Z_{{\bf B}L} (X, x_B) = R^{Z}_{0L} (X, x_B) \mathfrak{M}^{Z} (X) \label{Z0-reflection}
\eea
where $\mathfrak{M}^{Z}$ is given by
\be
\mathfrak{M}^{Z} (X) = \left({X^++{1\over X^+}\over
X^-+{1\over X^-}}\right)\left({X^++{1\over X^+}-{i\over g}\over
X^-+{1\over X^-}+{i\over g}}\right)
\mathfrak{M}^Y (X^\pm) \, .
\label{extra}
\ee
Again, analytic properties fit to our interpretation of the localized mode as $m=2$ magnon bound-state at the maximum momentum and relation of the amplitude to the bulk scattering amplitude via reflection-double. The reflection amplitude (\ref{Z0-boundstateR}) is essentially (\ref{decomposition}) times $\mathfrak{M}^Z(X)$. In relating the reflection-double of Fig. 2 with Fig. 1, we note that product of off-diagonal $S$ matrices is independent of $x_B$ and yields the reflection amplitude for $Y=0$ brane. According to our interpretation, product of diagonal $\mathcal{S}$ matrices gives bulk scattering amplitude for $m=2$ magnon bound-state and ${\bf B}^{(n)}$. Indeed, the first factor in (\ref{extra}) combined with the first factor of $\mathfrak{M}_{\B} (x_B, X)$ in (\ref{Z0-SB}) reproduces the double pole. The second factor in (\ref{extra}) corresponds to the $t$-channel pole. Finally, $A(x_B, X)$ times $\sigma^2 (X^\pm, x_B^\pm)$ corresponds to the two-body scattering amplitude. As such,
comparing reflection-double of Fig.2 with Fig. 1, we should
take square-root of Fig. 1 only for the off-diagonal contribution.

The factor $\widetilde{A}^{Z}_L$ in the reflection amplitude (\ref{Z0-boundstateR}) has a simple pole at
\be
x_B = X^-
\ee
with the energy
\bea
E_B ={i\over g}\left[ \left(X^+ -{1 \over X^+}\right)-\left(X^- -{1 \over X^-}\right)\right]
+{i\over g}\left(x_B -{1 \over x_B}\right)= \sqrt{(n+1)^2 + 4g^2}\,.
\eea
This pole corresponds to the excited state of the boundary degree formed by binding the $n$-magnon bound-state to the elementary boundary degree.
The remaining factors in $\mathfrak{M}^{Z}$ do not give rise to any new bound-state poles for the same reason as the $Y=0$ brane case.

\section{Strong Coupling Limit}

To confirm our results, we take the strong coupling limit and compare them with classical string worldsheet computations. The magnon bound-state is described as a soliton in complex sine-Gordon equation \cite{dorey}. The comparison was already made for dyonic giant magnon scattering in the bulk. In this limit, adopting the notation of \cite{AF}, the scattering phase-shift for the bound-state ${\bf B}$ takes the form
\bea
\delta_{\B} (X^\pm, Y^\pm) = 2g \left[ K(X^+, Y^+) + K(X^-, Y^-) - K(X^+, Y^-) - K(X^-, Y^+) \right]
\label{bulkphaseshift}
\eea
There are two sources contributing to $K(X, Y)$: the dressing phase-factor and the bound-state
phase-factor. From $- [\log  \sigma^2(X, Y)]/2g i$ of the dressing phase-factor, we extract that
\bea
K_{\rm dressing} (X, Y) = - \left[ \left( X+ {1 \over X} \right) - \left( Y + {1 \over Y} \right) \right] \log \left(1 - {1 \over XY} \right) \, . \label{dressing}
\eea
From$[\log \mathfrak{M}_{\bf B}(X, Y)]/2gi$ of the bound-state phase-factor, taking account of the giant magnon regime, we also extract that
\bea
K_{\rm bound-state} (X, Y) = \left[\left( X + {1 \over X} \right) - \left(Y + {1 \over Y} \right)\right] \log \left[ \left(X + {1 \over X} \right) - \left(Y + {1 \over Y} \right) \right]. \label{bound-state} \eea
In string worldsheet computations, the phase-shift was computed from time-delay in scattering two solitons of complex sine-Gordon model. For the bulk scattering, the two results were in complete agreement \cite{dorey}. We now want to check if the same holds for the reflection phase-shifts.

For $Y=0$ brane case, because (\ref{yzerobound}) is the square-root of the bulk scattering amplitude, the reflection phase-shift is immediately given by
\bea
\delta_L^{Y} (X^\pm) = {1 \over 2} \delta_{\B} (- X^\mp,\, X^\pm).
\label{Y0-phaseshift}\eea
In string worldsheet computations, the corresponding phase-shift is computable from the method of image. The time delay off the boundary equals to the half of the scattering between two identical solitons carrying opposite momenta. Therefore, the two results agree with each other.

For $Z=0$ brane case, the boundary mode contribution $\sigma^2(X^\pm, x_B^\pm)$ that enters through
$R_{0L}^Z(X, x_B)$ in (\ref{Z0-boundstateR}) adds extra shift to that common to $Y=0$ brane (\ref{Y0-phaseshift}). Quite remarkably, noting that $x_B^\pm \rightarrow \pm i$, we find that
$K_{\rm bound-state}(x_B, X) $ equals zero. It implies that, in the dyonic giant magnon regime, this contribution is universal for any kind of the boundary mode, elementary or composite. This leads to the conclusion that the total reflection phase-shift is given by
\bea
\delta_L^{Z} (X, x_B) = {1 \over 2} \delta_{\B} (- X^\mp,\, X^\pm)
+ \delta_{\B} (x_B,\, X^\pm) \, ,
\eea
where $x_B^\pm \rightarrow \pm i$. In string worldsheet computations, the second term (boundary mode contribution) admits an intuitive understanding: in the method of image, this phase-shift arises from scattering the soliton and its image soliton off a fixed soliton sitting at the boundary \cite{HM}. Once again, both results agree with each other.

\section*{Acknowledgement}
We thank participants of 2007 APCTP Focus Program ``Liouville, Integrability and Branes (4)'' for enlightening conversations. This work was supported in part by KRF-2007-313-C00150 (CA), ABRL R14-2003-012-01002-0 (DB), SRC-CQUeST-R11-2005-021 (DB, SJR), KRF-2005-084-C00003, EU FP6 Marie Curie Research \& Training Networks MRTN-CT-2004-512194 and HPRN-CT-2006-035863 through MOST/KICOS, and F.W. Bessel Award of Alexander von Humboldt Foundation (SJR).


\end{document}